\documentclass[twoside,reqno]{HERON}
\usepackage{epsfig,cite,colordvi}
\usepackage{graphicx}
\usepackage{url}
\usepackage{amssymb,amsmath,amscd,epsf}
\usepackage{times}
\usepackage{makeidx}
\usepackage{rotating}
\usepackage{multirow}
\usepackage{supertabular}

\begin{document}

\title{Coherent quadrupole-octupole states from a SUSY-QM Hamiltonian hierarchy
and shape invariance}

\runningheads{Coherent quadrupole-octupole states from a SUSY-QM Hamiltonian
hierarchy ...}{N. Minkov, S. Drenska, P. Yotov}

\begin{start}

\author{N. Minkov}{ }, \coauthor{S. Drenska}{ }, \coauthor{P. Yotov}{ }

\index{Minkov, N.} \index{Drenska, S.} \index{Yotov, P.}

\address{Institute of Nuclear Research and Nuclear Energy, Bulgarian
Academy of Sciences, Tzarigrad Road 72, BG-1784 Sofia, Bulgaria}{}

\begin{Abstract}

We show that the potential in the radial equation in the model of coherent
quadrupole-octupole motion (CQOM) in nuclei generates a sequence of superpotentials and
subsequent series of effective potentials which satisfy the shape-invariance condition and
correspond to a SUSY-QM hierarchy of Hamiltonians. On this basis we suggest that the
original CQOM level scheme possesses a generic supersymmetric structure of the spectrum
inherent for the coherent quadrupole-octupole mode. We outline the mechanism in which the
real quadru\-pole-octupole spectra in even-even and odd-even nuclei deviate from the
genuine symmetry. By using it we illustrate the possibilities to identify the signs of
supersymmetry in the alternating-parity spectra of even-even nuclei and the
quasi-parity-doublet levels of odd-mass nuclei described within the CQOM model approach.
\end{Abstract}
\end{start}
%\def\ref{\par\smallskip\hangindent=.6cm\hangafter=1}

%\parindent=0.pt

\section{Introduction}
\label{intro}

The supersymmetry (SUSY) concept in physics was introduced as a part of the efforts for a
unified description of the basic interactions in nature \cite{GRNWWZ}. Within SUSY one
may consider the fundamental particles -- fermions and bosons --  as superpartners  related by
a transformation which keeps the same mass and changes the spin by 1/2. Since the masses of
the presently observed particles do not allow one to identify any pair of superpartners, it
appears that the genuine SUSY does not exist at the currently accessible energies, whereas a
spontaneous symmetry breaking mechanism providing different masses for the superpartners
may take a place \cite{SUSYbr}. Nevertheless, the specific algebraic structure of SUSY,
which includes a combination of commutation (bosonic) and anti-commutation (fermionic)
relations, can be associated in a more general context with the solution of some quantum
mechanical problems and has lead to the development of the so-called supersymmetric
quantum mechanics (SUSY-QM) \cite{W81,CF83}. In particular, it was realized \cite{CKS}
that the SUSY-QM leads to a deeper understanding of the factorization approaches
\cite{Schr40,IH51} applied in the solution of the Schr\"odinger equation and allows one to
outline and properly systematize the various classes of known analytically solvable potentials.
A common feature of all these potentials, hereafter called SUSY potentials, is that each of
them generates a hierarchy of factorized Hamiltonians the eigenvalues of which  exhaust the
spectrum of the given potential. Also, it is known that some SUSY potentials satisfy the
so-called shape invariance condition  \cite{G83}. In this case all potentials in the hierarchy
have the same functional dependence on the (space) variable and only differ through a set of
discrete parameters which change under given rule at the subsequent hierarchy steps. The
different potentials are related through certain recurrence relation allowing one to obtain
simplified expressions for the energies and wave functions in the spectrum. A classification of
some basic shape invariant potentials (SIPs) is given in \cite{DKS}.

The SUSY-QM formalism was extended to Hamiltonians with coordinate-dependent effective
mass \cite{QT04,BBQT05}. The shape-invariance condition was generalized (deformed) for
the respectively  obtained effective potentials allowing one to apply the SUSY-QM techniques
for solving the eigenvalue problem. Recently this concept was applied to nuclear collective
models \cite{BGLMQ10,BGLMQ11,BGLMQ13}. The SUSY-QM technique for SIPs was
applied to solve the eigenvalue problem for Bohr-like Hamiltonians with
deformation-dependent mass terms in the cases of Davidson \cite{BGLMQ10,BGLMQ11} and
Kratzer \cite{BGLMQ13} potentials. The approach allows one to obtain analytical expressions
for spectra and wave functions for separable potentials in the cases of axially-symmetric
prolate deformed nuclei, $\gamma$-unstable nuclei  and triaxial nuclei. The dependence of
the mass on the deformation moderates the increase of the moment of inertia with
deformation, removing a known drawback of the Bohr model. As a result a good description
of ground-, $\beta$-, $\gamma$- energy band levels and the attendant B(E2) transition
probabilities in a wide range of nuclei in different regions of collectivity was obtained with a
reasonable accuracy.

Another collective model assuming a coherent quadrupole-octupole motion (CQOM) in
nuclei was developed by using a two-dimensional potential depending on the axial
quadrupole and octupole deformation variables \cite{MYDSBLP06}--\cite{MDDSLS13}. The
assumption of coherence allows one to exactly separate the variables in ellipsoidal coordinates
obtaining a ``radial'' equation for the effective quadrupole-octupole deformation and
``angular'' equation for the relative quad\-rupole-octupole excitation modes. The radial
equation involves a Davidson-like potential which allows one to find an analytical solution of
the model. The obtained spectrum corresponds to coherent quadrupole-octupole vibrations
coupled to rotation motions of the nucleus. It was shown that the  model is capable to
describe and classify the yrast and higher excited alternating-parity sequences in even-even
nuclei and split parity-doublet spectra in odd-mass nuclei. The analytical solvability and
classification ability of the model make it interesting to examine the possibly underlying
symmetry which determines the properties of the model system as well as to check to what
extent such a symmetry can be identified in the observed experimental spectra.

The aim of the present work is to clarify the above issue by applying the SUSY-QM
techniques to the  CQOM model Hamiltonian. It will be shown that the potential in the radial
equation is shape invariant and this allows one to express the Hamiltonian in terms of the
SUSY hierarchy and to subsequently obtain the model spectrum through the SUSY-QM
procedure. We shall examine the possibility to associate SUSY with the coherent
quadrupole-octupole modes assumed in the model. At the same time we shall study the
deviation of the experimentally observed quadrupole-octupole spectra from the model
imposed scheme. This should allow us to look for a proper symmetry-breaking mechanism
which may govern the observed collective properties of nuclei.

In Sec.~\ref{sec:2} an overview of the SUSY-QM and SIP concepts is given by using the
formulations provided in \cite{CKS}. In Sec.~\ref{sec:3} the application of the SUSY-QM
techniques to the CQOM model is presented. In Sec.~\ref{sec:4} the possibility to associate
CQOM with SUSY is considered together with the analysis of observed spectra and related
discussion. In Sec.~\ref{sec:5} concluding remarks are given.

\section{SUSY-QM formalism for analytically solvable potentials}
\label{sec:2}

\subsection{Supersymmetric factorization of a solvable one-dimensional Hamiltonian}

Let us consider the ground-state (gs) wave function $\psi_0(x)$ and energy $E_0$ for a
one-dimensional potential $V(x)$ which satisfy the Schr\"{o}dinger equation
\begin{eqnarray}
H\psi_0(x)=-\frac{\hbar^2}{2m}\frac{d^2\psi_0(x)}{dx^2}+V(x)\psi_0(x)=E_{0}\psi_0(x).
\label{Schrn}
\end{eqnarray}
If $\psi_0(x)$ is nodeless and the potential $V(x)\rightarrow V_{-}(x)=V(x)-E_0$ is shifted
so as $E_{0}\rightarrow E^{-}_0=0$, with $\psi_0^{-}(x)\equiv\psi_0(x)$, one obtains from
(\ref{Schrn})
\begin{eqnarray}
V_{-}(x)=\frac{\hbar^2}{2m}\frac{\psi_0^{''-}(x)}{\psi_0^{-}(x)}\ ,
\label{Vminus}
\end{eqnarray}
and  $H\rightarrow H_{-}$ such that
\begin{eqnarray}
H_{-}\psi_0^{-}(x)=-\frac{\hbar^2}{2m}\frac{d^2\psi_0^{-}(x)}{dx^2}+V_{-}(x)\psi_0^{-}(x)=0.
\label{Schr}
\end{eqnarray}
If the function $\psi_0^{-}(x)$ is known or guessed the potential $V_{-}(x)$ can be
determined from (\ref{Vminus}) up to a constant. Hamiltonian (\ref{Schr}) can be factorized
in the form \cite{CKS}
\begin{eqnarray}
H_{-}=A^{\dag}A. \label{factmin}
\end{eqnarray}
The operators $A$ and $A^{\dag}$ are defined as first order differential operators
\begin{eqnarray}
A=\frac{\hbar}{\sqrt{2m}}\frac{d}{dx}+W(x),\ \
A^{\dag}=-\frac{\hbar}{\sqrt{2m}}\frac{d}{dx}+W(x),\label{A}
\end{eqnarray}
where the unknown function $W(x)$ is determined so that Eq.~(\ref{Schr}) is satisfied after
introducing (\ref{factmin}) and (\ref{A})
\begin{eqnarray}
H_{-}\psi_0^{-}(x)=
\left[-\frac{h^2}{2m}\frac{d^2}{dx^2}-\frac{h}{\sqrt{2m}}W'(x)+W^{2}(x)\right]\psi_0^{-}(x)=0.
\label{SchrW}
\end{eqnarray}
By comparing Eqs. (\ref{SchrW}) and (\ref {Schr}) one finds:
\begin{eqnarray}
V_{-}(x)=W^2(x)-\frac{\hbar}{\sqrt{2m}}W'(x). \label{Vmin}
\end{eqnarray}
This equation is a first order differential equation for the function $W(x)$ known as Riccati
equation \cite{Riccati}. By taking $V_{-}(x)$ from  (\ref{Vminus}) its solution is obtained as
\cite{CKS}
\begin{eqnarray}
W(x)=-\frac{\hbar}{\sqrt{2m}}\frac{\psi_0^{'-}(x)}{\psi_0^{-}(x)}=
-\frac{\hbar}{\sqrt{2m}}\frac{d}{dx}\ln[\psi_0^{-}(x)].
\label{W}
\end{eqnarray}
Then the action of $A$ on the gs wave function gives
\begin{eqnarray}
A\psi_0^{-}(x)=\frac{\hbar}{\sqrt{2m}}\frac{d}{dx}\psi_0^{-}(x)+W(x)\psi_0^{-}(x)=0,
\label{Apsi0}
\end{eqnarray}
which allows one to determine $\psi_0^{-}(x)$  by $W(x)$
\begin{eqnarray}
\psi_0^{-}(x)=N\exp\left(-\frac{\sqrt{2m}}{\hbar}\int^{x}W(k)dk\right)\ ,
\label{grwave funct}
\end{eqnarray}
where $N$ is a normalization constant.

By changing the order of the operators  in the factorization (\ref{factmin}) one defines a new
Hamiltonian
\begin{eqnarray}
H_{+}=AA^{\dag}, \label{factplus}
\end{eqnarray}
whose action on the gs wave function
\begin{eqnarray}
H_{+}\psi_0^{-}(x)=
\left[-\frac{h^2}{2m}\frac{d^2}{dx^2}+\frac{h}{\sqrt{2m}}W'(x)+W^{2}(x)\right]\psi_0^{-}(x)
\end{eqnarray}
defines a new potential \cite{CKS}
\begin{eqnarray}
V_{+}(x)=W^2(x)+\frac{\hbar}{\sqrt{2m}}W'(x)
\label{Vpl}
\end{eqnarray}
related to $V_{-}(x)$, $W(x)$ and $W'(x)$  through
\begin{eqnarray}
\frac{1}{2}[V_{-}(x)+V_{+}(x)]=W^{2}(x), \ \
\frac{1}{2}[V_{-}(x)-V_{+}(x)]=W'(x) \ .
\end{eqnarray}
Also, one has
\begin{eqnarray}
[A,A^{\dag}]=\frac{-2\hbar}{\sqrt{2m}}W'(x).
\end{eqnarray}
The function  $W(x)$ is known as {\em superpotential}, while the potentials $V_{-}(x)$ and
$V_{+}(x)$ are called {\em supersymmetric partners}.

By considering the eigenfunctions $\psi_n^{-}(x)$, $\psi_n^{+}(x)$ and eigenvalues
$E_{n}^{-}$, $E_{n}^{+}$ ($n=0,1,2\dots$) of the Hamiltonians $H_{-}$ and $H_{+}$,
respectively, one can easily check that
$H_{+}(A\psi_{n}^{-}(x))=E_{n}^{-}(A\psi_{n}^{-}(x))$ and
$H_{-}(A^{\dag}\psi_{n}^{+}(x))=E_{n}^{+}(A^{\dag}\psi_{n}^{+}(x))$ which together
with (\ref{Apsi0}) leads to the relations
\begin{eqnarray}
E_{n}^{+}&=&E_{n+1}^{-}\ , \ \ \ \mbox{with } \  E_{0}^{-}=0\\
\psi_{n}^{+}(x)&=&\left(E_{n+1}^{-}\right)^{-\frac{1}{2}}A\psi_{n+1}^{-}(x)
\ , \ \ \  \psi_{n+1}^{-}(x)=\left(E_{n}^{+}\right)^{-\frac{1}{2}}A^{\dag}\psi_{n}^{+}(x).
\label{psipm}
\end{eqnarray}
It is seen that the spectra of  $H_{-}$ and $H_{+}$ are identical except for $E_{0}^{-}$
which does not appear for $H_{+}$. By using (\ref{psipm}) one can obtain the
eigenfunctions of $H_{-}$ from those of  $H_{+}$ and vice versa, except for $\psi_0^{-}$
which is determined by the superpotential in (\ref{grwave funct}). $H_{-}$ and $H_{+}$ are
referred to as SUSY partners. Together with the operators $A$ and $A^{\dag}$ they form a
set of matrices \cite{CKS}
\[H= \left( \begin{array}{cc}
H_{-} & 0 \\
0 &   H_{+} \\
 \end{array} \right)\ \ \ \ \
Q= \left( \begin{array}{cc}
0 & 0 \\
A & 0  \\
 \end{array} \right)\ \ \ \ \ Q^{+}= \left( \begin{array}{cc}
0 & A^{\dag} \\
0 & 0     \\
 \end{array} \right),\]
which satisfy the following commutation and anti-commutation relations
\begin{eqnarray}
[H,Q]=[H,Q^{+}]=0,\ \ \
\{Q,Q^{+}\}=H, \ \ \ \{Q^{+},Q^{+}\}=\{Q,Q\}=0
\end{eqnarray}
closing the superalgebra of $sl(1,1)$. The operators $Q$ and $Q^{+}$ are known in the
SUSY theory as {\em supercharges} and can be interpreted as operators transforming the
bosonic degrees of freedom into fermionic ones. The fact that they commute with $H$ is
related to the supersymmetric degeneracy of the spectrum. The SUSY algebra is an extension
of the Poincare algebra \cite{W81,CKS}.

\subsection {SUSY-QM hierarchy of Hamiltonians and shape invariant potentials}

Starting by  $H^{(1)}\equiv H_{-}$ and $H^{(2)}\equiv H_{+}$, Eqs.~(\ref{factmin}) and
(\ref{factplus}), a {\em hierarchy of supersymmetric partner Hamiltonians} $H^{(1)}, \
H^{(2)}, H^{(3)},\dots$  can be constructed, so that $H^{(3)}$ is obtained as the upper
superpartner of $H^{(2)}$ after factorizing the latter by analogy to $H^{(1)}$, $H^{(4)}$ is
constructed by $H^{(3)}$ in the same way, and so on. In this procedure the levels of each
subsequent Hamiltonian are obtained from the previous one by removing its lowest level.
Thus, knowing the eigenvalues and eigenfunctions of $H^{(1)}$ one obtains the eigenvalues
and eigenfunctions of all Hamiltonians in the hierarchy. The energy levels appearing in the
hierarchy are illustrated schematically below.
%\vspace{-0.8cm}
\begin{eqnarray}
\label{eq2:}\begin{aligned}
 E_5^{(1)}   \frac{\hspace*{5mm}}{\hspace*{5mm}}& \;  & E_4^{(2)} \frac{\hspace*{5mm}}{\hspace*{5mm}}& \;
  & E_3^{(3)}   \frac{\hspace*{5mm}}{\hspace*{5mm}}& \;  & E_2^{(4)} \frac{\hspace*{5mm}}{\hspace*{5mm}}& \;
&  E_1^{(5)}   \frac{\hspace*{5mm}}{\hspace*{5mm}}& \; & E_0^{(6)} \frac{\hspace*{5mm}}{\hspace*{5mm}}& \; \\
  E_4^{(1)}   \frac{\hspace*{5mm}}{\hspace*{5mm}}& \;  & E_3^{(2)} \frac{\hspace*{5mm}}{\hspace*{5mm}}& \;
&  E_2^{(3)}    \frac{\hspace*{5mm}}{\hspace*{5mm}}& \;  & E_1^{(4)}
\frac{\hspace*{5mm}}{\hspace*{5mm}}& \;  & E_0^{(5)}
   \frac{\hspace*{5mm}}{\hspace*{5mm}}& \; \\
  E_3^{(1)}   \frac{\hspace*{5mm}}{\hspace*{5mm}}& \;   & E_2^{(2)} \frac{\hspace*{5mm}}{\hspace*{5mm}}& \;
&   E_1^{(3)}    \frac{\hspace*{5mm}}{\hspace*{5mm}}& \;  & E_0^{(4)} \frac{\hspace*{5mm}}{\hspace*{5mm}}& \; \\
  E_2^{(1)}  \frac{\hspace*{5mm}}{\hspace*{5mm}}& \;   & E_1^{(2)} \frac{\hspace*{5mm}}{\hspace*{5mm}}& \; & E_0^{(3)}
   \frac{\hspace*{5mm}}{\hspace*{5mm}}& \; \\
  E_1^{(1)}   \frac{\hspace*{5mm}}{\hspace*{5mm}}& \;  & E_0^{(2)} \frac{\hspace*{5mm}}{\hspace*{5mm}}& \; \\
  E_0^{(1)}   \frac{\hspace*{5mm}}{\hspace*{5mm}}& \; \\
  H^{(1)} & & H^{(2)}  & & H^{(3)}  &
&  H^{(4)}  & & H^{(5)}  & & H^{(6)}
\end{aligned}
\nonumber
\end{eqnarray}

%\subsection {Shape invariant potentials}

The pair of supersymmetric partner potentials $V_{\pm}(x)$ defined by (\ref {Vmin}) and
(\ref{Vpl})  represents a {\em shape invariant} if they  satisfy the condition\cite{G83}
\begin{eqnarray}
V_{+}(x;a_{1})=V_{-}(x;a_{2})+R(a_{1}), \label{shape inv cond}
\end{eqnarray}
where $a_{1}$ is a set of parameters, while $a_{2}=f(a_{1})$ and $R(a_{1})$ are functions
of $a_{1}$ but not of $x$. In this case all supersymmetric partner potentials appearing in the
hierarchy of Hamiltonians can be simply expressed in the form of $V_{-}$, so that
\begin{equation*}\label{eq3:}\begin{aligned}
H^{(1)} &=-\frac{h^2}{2m}\frac{d^2}{dx^2}+V_{-}(x;a_{1})   \\
H^{(2)} &=-\frac{h^2}{2m}\frac{d^2}{dx^2}+V_{-}(x;a_{2})+R(a_{1})  \\
... & ..........................................   \\
H^{(m)} &\equiv-\frac{h^2}{2m}\frac{d^2}{dx^2}+V_{-}(x;a_{m})+\sum_{k=1}^{m-1}R(a_{k}),
\end{aligned}\end{equation*}
where $a_{m}=f^{(m-1)}(a_{1})$. The gs energies of the different Hamiltonians are
\begin{eqnarray}
E_{0}^{(m)}=\sum_{k=1}^{m-1}R(a_{k})+E_{0}^{(1)}.
\end{eqnarray}
The complete energy spectrum of $H^{(1)}$ is
\begin{eqnarray}
E_{n}^{(1)  }=\sum_{k=1}^{n}R(a_{k}), \  \mbox{with} \ E_{0}^{(1)}=0, \ n=0,1,2,...
\label{energy}
\end{eqnarray}
while the corresponding eigenfunctions are determined by
\begin{eqnarray}
\psi_{n}^{(1)}(x;a_{1})\backsim
A^{\dag}(x;a_{1})A^{\dag}(x;a_{2})...A^{\dag}(x;a_{n})\psi_{0}^{(1)}(x;a_{n+1}),
\label{gwf}
\end{eqnarray}
where the operators $A^{\dag}(x;a_{n})$ are determined by (\ref{A}) with $W(x;a_{n})$
being a function of $V_{-}(x;a_{n})$ and $\sum_{k=1}^{n-1}R(a_{k})$.

\section{SUSY-QM formalism in the CQOM model}
\label{sec:3}

\subsection{The problem of soft quadrupole-octupole vibrations and rotations}

The Hamiltonian of quadrupole-octupole vibrations and rotations has been taken in the form
\cite{MYDSBLP06}--\cite{MDDSLS13}
\begin{eqnarray}
H_{\mbox{\scriptsize qo}}
&=&-\frac{\hbar^2}{2B_2}\frac{\partial^2}{\partial\beta_2^2}
-\frac{\hbar^2}{2B_3}\frac{\partial^2}{\partial\beta_3^2}+
U(\beta_2,\beta_3,I) ,\label{CQOMH}
\end{eqnarray}
where $\beta_{2}$ and $\beta_{3}$ are axial quadrupole and octupole variables and the
potential is
\begin{eqnarray}
U(\beta_2,\beta_3, I)=\frac{1}{2}C_2{\beta_2}^{2}+
\frac{1}{2}C_3{\beta_3}^{2} + \frac{X(I)}
{d_2\beta_2^2+d_3\beta_3^2}.
\end{eqnarray}
Here $B_2$ $(B_3)$, $C_2$ $(C_3)$ and $d_2$ ($d_3$) are quadrupole (octupole) mass,
stiffness and inertia parameters, respectively, and $X(I)$ involves the angular momentum
dependence of the spectrum specified in \cite{MYDSBLP06,MDYLBS07}. Under the
assumption of coherent quadrupole-octupole oscillations with a frequency
$\omega=\sqrt{C_2/B_2}=\sqrt{C_3/B_3}\equiv \sqrt{C/B}$ and by introducing ellipsoidal
coordinates $\beta_2=\sqrt{d/d_2}\eta$ $\cos\phi$, $\beta_3=\sqrt{d/d_3}\eta \sin\phi $, with
$d=(d_2+d_3)/2$, one obtains the Schr\"{o}dinger equation for (\ref{CQOMH}) with
separated variables
\begin{eqnarray}
\frac{d^2\psi(\eta)}{d\eta^2}+\frac{1}{\eta}\frac{d\psi}{d\eta}+
\frac{2B}{\hbar^2}\left[E-\frac{\hbar^2}{2B}\frac{k^2}{\eta^2}
-\frac{1}{2}C\eta^2-\frac{ X(I)}{d\eta^2}\right]\psi(\eta)=0
\end{eqnarray}
\begin{eqnarray}
\frac{\partial^2}{\partial \phi^2}\varphi(\phi)+k^2\varphi(\phi)=0 \ .
\label{angwf}
\end{eqnarray}
By substituting
\begin{eqnarray}
\psi(\eta)=\eta^{-1/2}\varphi(\eta)
\end{eqnarray}
the radial equation is obtained in the form
\begin{eqnarray}
\frac{d^2\varphi(\eta)}{d\eta^2}+\frac{2B}{\hbar^2}\left[E-\frac{\hbar^2}{2B}\frac{k^2}{\eta^2}
-\frac{1}{2}C\eta^2-\frac{X(I)}{d\eta^2}+\frac{1}{4}\frac{\hbar^2}{2B\eta^2}\right]\varphi(\eta)=0.
\label{CQOMrad}
\end{eqnarray}

\subsection{Shape invariance of the CQOM potential}

Applying the SUSY-QM procedure  to Eq.~(\ref{CQOMrad}), we consider the effective
potential in the ground state ($E=E_0$) as the first (lowest) superpartner potential
\begin{eqnarray}
V_{-}(\eta)=V_{\mbox{\scriptsize eff}}=\frac{1}{2}B\omega^{2}\eta^{2}+
\frac{\hbar^2}{2B\eta^{2}}\left (s^{2}-\frac{1}{4}\right )-E_0,
\label{Vmeta}
\end{eqnarray}
where $s=s(k,I)=\sqrt{k^2+bX(I)}$, with $b=2B/\hbar^2d$. (Note that in Refs.
\cite{MYDSBLP06}--\cite{MDDSLS13} the quantity $s$ includes an additional factor 1/2 ).
To find the SUSY partner potential $V_{+}$, Eq.~(\ref{Vpl}), the Riccati equation
(\ref{Vmin}) for the superpotential $W(\eta)$ has to be solved. We search $W(\eta)$ in the
form
\begin{eqnarray}
W(\eta)=a_1\eta-\frac{a_2}{\eta},
\label{Weta}
\end{eqnarray}
where $a_1$ and $a_2$ are parameters to be determined. By substituting (\ref{Vmeta}) and
(\ref{Weta}) into the left and right hand sides of (\ref{Vmin}), respectively, and after
equating the powers of $\eta$ in both sides one finds
\begin{eqnarray}
a_{1}=\omega\sqrt{\frac{B}{2}}, \ \ \ a_{2}=\frac{\hbar}{
\sqrt{2B}}(s+1/2), \ \ \ E_{0}(s)=\hbar \omega(1+s)\ .
\end{eqnarray}
As a result the superpotential $W(\eta)$ becomes
\begin{eqnarray}
W(\eta)=W(\eta,s)=\sqrt{\frac{B}{2}}\omega\eta-\frac{\hbar}{\sqrt{2B}}\frac{(s+1/2)}{\eta}\ ,
\label{Wetas}
\end{eqnarray}
and the potentials  $V_{-}(\eta,s)$ and $V_{+}(\eta,s)$ which now determine the partner
Hamiltonians $H^{(1)}$ and $H^{(2)}$ are obtained in the form
\begin{eqnarray}
V_{-}(\eta;s)&=&\frac{1}{2}B\omega^{2}\eta^{2}+\frac{\hbar^2}{2B\eta^{2}}(
s-1/2)(s+1/2) -\hbar \omega (s+1) \\
V_{+}(\eta;s)&=&\frac{1}{2}B\omega^{2}\eta^{2}+\frac{\hbar^2}{2B\eta^{2}}(
s+1/2)(s+3/2) -\hbar \omega s \ .
\end{eqnarray}
It is easily seen that $V_{-}$ and $V_{+}$ satisfy the shape invariance condition
Eq.~(\ref{shape inv cond}) with $R(s)=R=2\hbar \omega$ not depending on $s$.

\subsection{SUSY-QM hierarchy and solution of the CQOM eigenproblem}

By continuing the above procedure one gets the SUSY-CQOM potential partners in the
following general form
\begin{eqnarray}
V_{-}(\eta;s+m-1)&=&\frac{1}{2}B\omega^{2}\eta^{2}
+\frac{\hbar^2}{2B\eta^{2}}(s-\frac{1}{2}+m-1)(s+\frac{1}{2}+m-1)\nonumber \\
&+&2(m-1)\hbar \omega,  \nonumber \\   \nonumber \\
V_{-}(\eta;s+m)&=&\frac{1}{2}B\omega^{2}\eta^{2}+
\frac{\hbar^2}{2B\eta^{2}}(s-\frac{1}{2}+m)(s+\frac{1}{2}+m)\nonumber\\
&+&2m\hbar \omega.  \nonumber
\end{eqnarray}

As a result the SUSY-hierarchy Hamiltonians and their ground-state energies are obtained in
the following schematic form
\vspace{1.5cm}

$H^{(1)}\rightarrow V_{-}(s)+E_{0}(s) \rightarrow  E_{0}^{(1)}(s) $\\

$H^{(2)}\rightarrow V_{-}(s+1)+2 \hbar \omega +E_{0}(s)  \rightarrow E_{0}^{(2)}(s)= E_{1}^{(1)}(s)$\\

$H^{(3)}\rightarrow V_{-}(s+2)+4 \hbar \omega +E_{0}(s) \rightarrow  E_{0}^{(3)}(s)= E_{2}^{(1)}(s)$  \\

................................................................................\\

$H^{(m)}\rightarrow V_{-}(s+m-1)+2(m-1) \hbar \omega +E_{0}(s) \rightarrow E_{0}^{(m)}(s)= E_{m-1}^{(1)}(s)$\\

$H^{(m+1)}\rightarrow V_{-}(s+m)+2m \hbar \omega +E_{0}(s) \rightarrow
E_{0}^{(m+1)}(s)= E_{m}^{(1)}(s)$\ .
\medskip

According to  (\ref{energy}) the energies  $E_{n}(s)\equiv E_{n}^{(1)}(s)$, $n=0,1,2,\dots$
form the full spectrum of $H^{(1)}$ which enters the radial CQOM equation
(\ref{CQOMrad}). Thus one obtains the CQOM spectrum given in \cite{MYDSBLP06}
\begin{eqnarray}
 E_{n}(s)= E_{nk}(I)= \hbar \omega\left [ 2n+s(k,I)+1\right ]
 =\hbar\omega \left[ 2n+1+\sqrt{k^2+bX(I)}\right].
 \label{CQOMen}
\end{eqnarray}

The ground state wave function for $H^{(1)}$, with $n=0$, is obtained by introducing
$W(\eta,s)$, Eq~(\ref{Wetas}), into Eq.~(\ref{grwave funct})
\begin{eqnarray}
f_{0}(\eta ,s) &=&N_{0}\exp\left[-\frac{\sqrt{2B}}{\hbar}\int^{\eta}\left(\sqrt{\frac{B}{2}}\omega
x-\frac{\hbar}{\sqrt{2B}}\frac{(s+1/2)}{x}\right) dx\right]\nonumber \\
&=&N_{0}\, \exp\left(-\frac{c \eta ^{2}}{2}\right)\eta^{(s+1/2)}, \hspace{0.3cm}
\label{susyf0}
\end{eqnarray}
where $c=\sqrt{BC}/\hbar$ and the normalization factor is $N_{0}=\sqrt{
\frac{2c^{s+1}}{\Gamma (s+1)}}$. Further, by applying the general expression (\ref{gwf})
for the next state ($n=1$) one finds
\begin{eqnarray}
f_{1}(\eta , s)&=&N_{1} A^{\dag}(\eta ,s)f_{0}(\eta ,s+1) \nonumber \\
&=&N_{1}(-c\eta^{2}+s+1)\exp\left(-\frac{c \eta
^{2}}{2}\right)\eta^{(s+1/2)}\ , \label{susyf1}
\end{eqnarray}
with  $N_{1}=\sqrt{ \frac{2c^{s+1}}{\Gamma (s+2)}}$.  By continuing this procedure, one
finds by induction the known radial wave function of CQOM \cite{MDSSL12}
\begin{equation}
f_{n}(\eta , s)=\psi^I_{n,k}(\eta )=\sqrt
{\frac {2c\Gamma(n+1)}{\Gamma(n+2s+1)}}
e^{-c\eta^2/2}(c\eta^{2})^sL^{2s}_n(c\eta^2)\ , \label{psieta1}
\end{equation}
which involves the generalized Laguerre polynomials in the variable $\eta$.

\section{The meaning of SUSY and its breaking in CQOM: Discussion}
\label{sec:4}

\begin{figure}[t]
\centering \centerline{\includegraphics[scale=0.39]{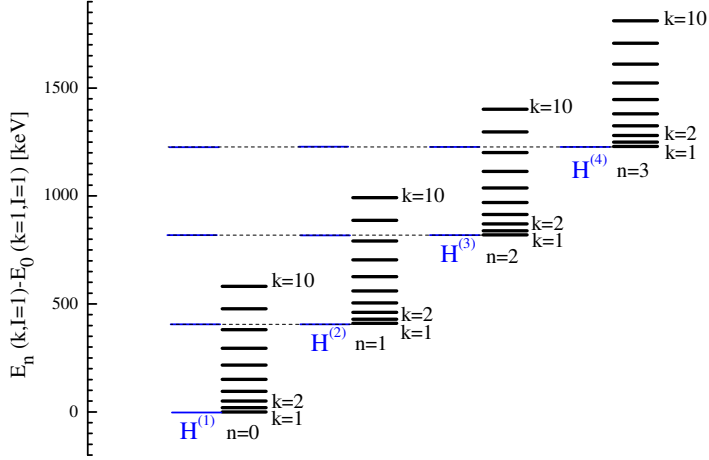}}
\caption[]{Schematic quadrupole-octupole vibration levels (without rotations) and
SUSY-QM hierarchy (in blue) of the bandhead states in the CQOM model.}
\label{f1}
\end{figure}

The quadrupole-octupole vibration spectrum obtained by the analytic expression
(\ref{CQOMen}) is schematically illustrated in Fig.~\ref{f1}. It consists of $k=1,2,3\dots$
level-sequences built on different $n=0,1,2,\dots$ excitations. In addition, on each $k$-level a
rotation band (not given for simplicity) is built, with the allowed angular momentum values
depending on parity conditions imposed by the model (see below). The different
$n$-sequences have identical structures and are equally shifted by $2\hbar\omega$. It is seen
that the bandheads $(n=0,1,2,\dots, k=1, I=I_{\mbox{\scriptsize bh}})$ of these sequences
generate the full SUSY hierarchy of levels (given in Fig.~\ref{f1} in blue) which appear in the
CQOM scheme.  Here $I_{\mbox{\scriptsize bh}}$ is the lowest possible (bandhead) angular
momentum for given $n$, which is $0$ for the even-even nuclei and $1/2$ for odd-mass
nuclei. The set of bandhead levels $E_{n,k=1}(I_{\mbox{\scriptsize bh}})$ determines the
SUSY-QM content of the CQOM model. It suggests that the assumed coherent
quadrupole-octupole  motion in nuclei possesses a {\em genuine SUSY inherent for the radial
($\eta$) vibration mode}. At the same time the development of the $k$-sequences (i.e. the
angular ($\phi$) mode) and the further superposed rotation levels can be interpreted as the
result of a {\em dynamical breaking of SUSY} in which the full spectrum of the system is
generated. In the present context the term ``dynamical'' means that the symmetry breaking is
due to the involvement of additional dynamic modes (angular vibrations and rotations)
represented by the quantum numbers $k$ and $I$ outside of the one-dimensional radial
motion.
%Still the so determined model spectrum is too schematic and idealized.

\begin{figure}[t]
\centering \centerline{\includegraphics[scale=0.31]{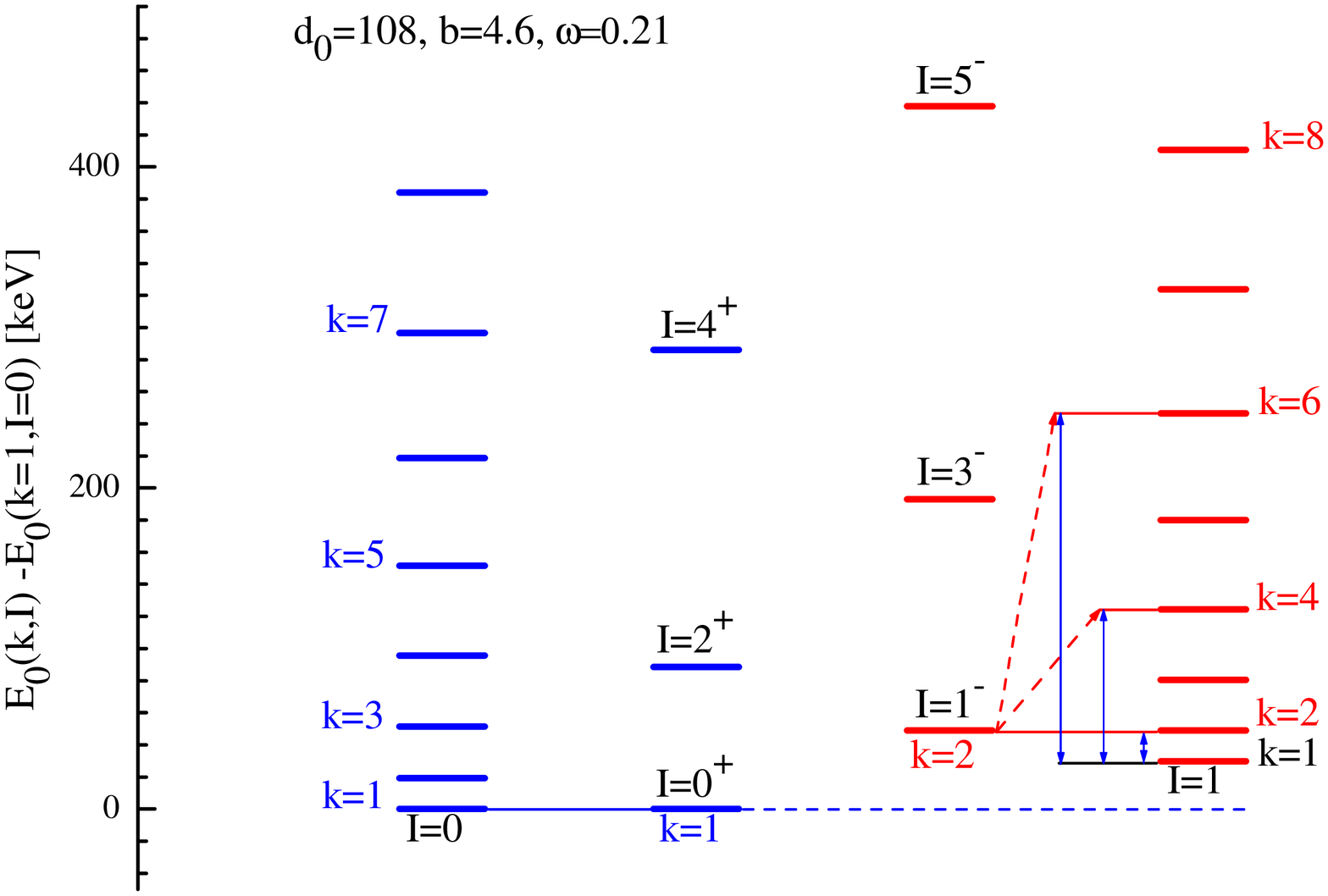}}
\ \
\centerline{\includegraphics[scale=0.32]{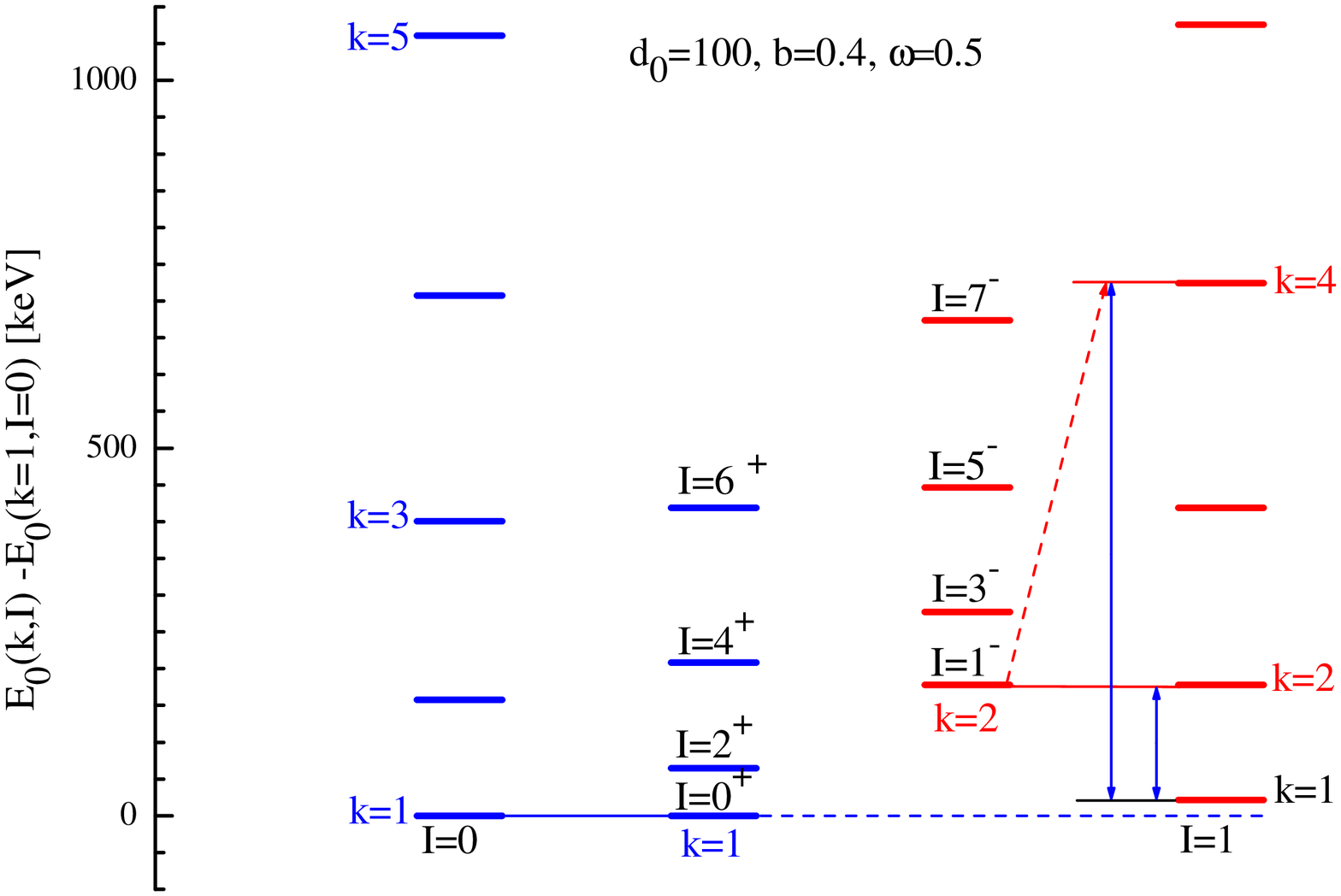}}
%\centering \centerline{\includegraphics[scale=0.45]{cqom_1minus_up.eps}}
\caption[]{Schematic structure of alternating-parity spectrum in CQOM with small (upper part)
and large (lower part) parity shift. The energy levels are obtained by (\ref{CQOMen}) with
$X(I)=[d_0+I(I+1)]/2$, where $d_0$ is a parameter of the potential.}
\label{f2}
\end{figure}

Further, by applying the model scheme to the quadrupole-octupole spectra one imposes
geometrically motivated parity conditions. In the alternating-parity bands of even-even nuclei
the positive-parity states correspond to an odd $k$-value, while the negative-parity states
correspond to even $k$ \cite{MYDSBLP06,MDSSL12}. In the quasi-parity-doublets of
odd-mass nuclei this correspondence depends on the odd-particle parity
\cite{MDYLBS07,MDDSLS13}. The difference in the two $k$-values generates the
parity-shift effect.  This is illustrated for the alternating-parity bands in Fig.~\ref{f2} by taking
the lowest possible values $k=1$ and $2$. The magnitude of the parity shift can be adjusted
by changing the parameters values (especially $b$ and/or $\omega$), as seen by comparing
the spectra in the upper and lower parts of Fig.~\ref{f2}. (The shift is indicated by blue
double-arrows). However, if the parity-shift differs for the different sequences in given
spectrum, one needs to consider different pairs of $k$-values, like $(1,2)$, $(1,4)$, $(1,6)$ as
shown in Fig.~\ref{f2} or others (as e.g. in \cite{MDSSL12}), in order to reproduce the
observed energy displacements. This means that a deeper mechanism of symmetry breaking
may take place when the CQOM model scheme is applied to reproduce the realistic
quadrupole-octupole spectra.

Now we are able to examine to what extent the so-defined broken SUSY manifests in the
experimental spectra of nuclei with quadrupole-octupole degrees of freedom. We can say that
if these spectra possess it as a {\em generic symmetry} one should observe different sets of
(almost) identical band-structures -- alternating-parity bands in even-even nuclei and
quasi-parity-doublets in odd-mass nuclei -- shifted one from another by (almost) the same
energy intervals. Indeed by looking into data one finds several examples where such structure
of the spectrum can be identified. Good examples are observed for the nuclei $^{152}$Sm,
$^{154}$Gd and $^{100}$Mo whose CQOM-theoretical and experimental spectra are given in
Figs. 1, 3 and 7, respectively of Ref.~\cite{MDSSL12}. In the three nuclei it is seen that the
structures of the yrast and non-yrast alternating-parity sequences are very similar and in
addition in $^{154}$Gd the spacing between the three $0^{+}$ bandheads is almost the same.
Reasonable examples in odd-mass nuclei are the spectra of  $^{223}$Ra (Fig.~2 in
Ref.~\cite{MDDSLS13}) and $^{237}$U (Fig.~4 in Ref.~\cite{MDDSLS12}).

From another side a wider look on experimental data shows the presence of
quadrupole-octupole spectra in which the different energy sequences are neither quite
identical nor really equidistantly displaced. Such are the spectra in $^{154}$Sm, $^{156}$Gd,
$^{158}$Gd and $^{236}$U (Figs. 2, 4, 5 and 6 in Ref.~\cite{MDSSL12}) as well as
$^{161}$Dy and $^{239}$Pu  (Figs. 2 and 4 in Ref.~\cite{MDDSLS11}). In these cases,
especially in even-even nuclei, the different band structures (including parity-shifts) are
reproduced by the model through introducing quite different pairs of $k$-values, whereas the
energy displacements are reproduced by taking the lowest $k$ larger than 1.  We should
remark that in odd-mass nuclei the symmetry is additionally violated by the odd nucleon
which not only affects the even-even core through the parity and Coriolis effects, but also
may add a quasiparticle excitation energy to the doublet bandheads.

\section{Conclusion}
\label{sec:5}

We have shown that the CQOM formalism, in which the potential is a shape invariant in the
space of the effective quadrupole-octupole deformation, can be interpret in terms of the
SUSY-QM approach. As a result the radial quadrupole-octupole vibrations of the nucleus are
associated with a SUSY hierarchy of Hamiltonians, which together with the angular and
rotation modes provide the full model spectrum of the system. This suggests that a nucleus
capable of performing coherent quadrupole-octupole motions may posses a genuine SUSY,
which determines the basic structure of its collective excitation spectrum. The considered
examples show that if present in quadrupole-octupole spectra, the SUSY should be
necessarily and multilaterally broken in dependence on various conditions and particular
structure effects inherent for the even-even and odd-mass nuclei. Nevertheless, the
application of the SUSY-QM concept to the CQOM model approach allows one to get a
deeper insight into the symmetry properties of complex deformed nuclei as well as to better
understand (and justify) the proposed model classification of nuclear quadrupole-octupole
excitations. This result supports the relevance of the model scheme in nuclear
quadrupole-octupole spectra and its applicability as a basis to solve the more general problem
of quadrupole-octupole motions beyond the restrictions imposed by the coherent mode.

\end{document}